# PAI Data

Summary of the Project PAI Data Protocol

Authors:
Jincheng Du[1]
Dan Fang, PhD[2]
Mark Harvilla, PhD[3]

## Abstract

The Project PAI Data Protocol ("PAI Data") is a specification that extends the Project PAI Blockchain Protocol to include a method of securing and provisioning access to arbitrary data. In the context of PAI Coin Development Proposal (PDP) 2 [1], this paper defines two important transaction types that PAI Data supports: Storage Transactions, which facilitate storage of data and proof of ownership, and Sharing Transactions, designed to enable granting and revocation of data access to designated recipients. A comparative analysis of PAI Data against similar blockchain-based file storage systems is also presented.

---

[1] Blockchain Researcher at ObEN, Inc.
[2] Blockchain Researcher at ObEN, Inc.
[3] Chief Engineer at ObEN, Inc.

# Introduction

The Project PAI Data Protocol ("PAI Data") is a specification that extends the Project PAI Blockchain Protocol to include a method of securing and provisioning access to arbitrary data. This is achieved by utilizing a custom OP_RETURN data protocol and elliptical encryption, in combination with a data store (*e.g.*, a segmented network of torrent nodes for decentralized storage or a proprietary data store for centralized storage) [1].

## PAI Data Background

PAI Data is at the core of Project PAI's effort to build a sustainable, humanistic information economy [2]. With PAI Data, users can secure their data, and permission such data, using the Project PAI Protocol. PAI Pass, an application currently in development, provides an illustration of potential use cases that leverage PAI Data. PAI Pass is an integrated single sign-on service and data management application. The PAI Pass platform enables users to input and secure their data (*e.g.*, name, age, etc.) using a user-chosen third party data store protocol independent of the PAI Pass platform such as PAI Data.

## Other Use Cases

While not a substitute for traditional methods for protection of intellectual property (*e.g.*, copyright registration), PAI Data could be extremely useful for developers of intellectual property who want to commercialize their content while preserving evidence of authorship. As described below, PAI Data provides a simple way to store data with an immutable timestamp that can be validated by a third party. As a result, the use of PAI Data to secure proprietary data (e.g., trade secrets and/or other intellectual property) together with the identity of the author or owner of such data, would result in strong evidence that, on or before the time reflected in the timestamp: (1) a work had come into existence; and (2) the author identified had created, or at the very least had knowledge of, such data. This functionality would be especially useful if augmented by an authentication system (*e.g.*, PAI Pass).

# Technical Description

The main components of the PAI Data protocol are:

- Project PAI blockchain
- OP_RETURN op code
- Elliptical Encryption
- Submitter + Provider + Recipient
- Data Store (*e.g.*, a segmented network of torrent nodes)

# Transaction Types

## Storage Transaction

A storage transaction is used to store data via the PAI Blockchain and, provided said transaction includes information about a party claiming ownership or is used in conjunction with a platform such as PAI Pass, can be used to make a claim of ownership of the stored data. The storage transaction is created by broadcasting a transaction to oneself, with an OP_RETURN output containing a hashed identifier of the data for retrieval via a Storage Provider. Alternatively, for example, the OP_RETURN may contain a hash of the data itself, or some other identifier, to be used for mapping to data in a proprietary data store.

## Sharing Transaction

A sharing transaction is used to share data between a sender and a recipient via the PAI Blockchain. The sharing transaction is created by broadcasting a transaction to an intended recipient, with an OP_RETURN output containing a hashed identifier of the data for retrieval via a Storage Provider.

# Use Cases

This section outlines how the PAI Coin protocol can be used depending on your needs.

| *If you want to...* | *You should use a...* |
| --- | --- |
| Store and/or claim ownership of data (*e.g.*, digital assets, intellectual property) | PAI Storage Transaction |
| Permission data to a recipient | PAI Sharing Transaction (with a "grant" Operation ID, according to PDP 2 [1]) |
| Revoke recipient's permission to use data | PAI Sharing Transaction (with a "revoke" Operation ID, according to PDP 2 [1]). |

## Example: Storage Transaction

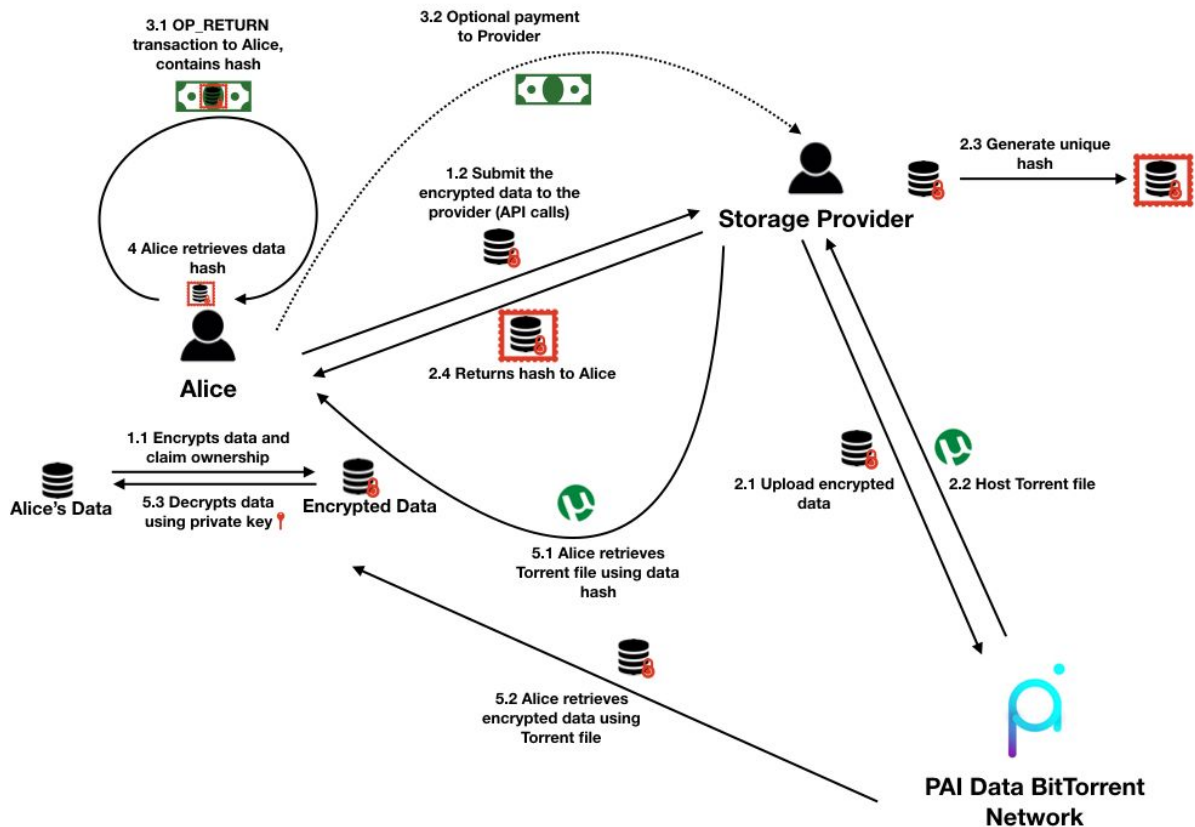

*Figure 1. Storage Transaction Workflow*

As illustrated in Figure 1, a party (e.g., Alice) can store data and claim ownership of the stored data via PAI Data transaction, in the following way.

1. Alice encrypts the data, along with a claim of ownership or authorship, with her public key. This encrypted data can only be decrypted by Alice, who has the corresponding private key.
2. Alice chooses a storage provider and, through a series of standard API calls specified by the protocol (i.e., the "Provider API"), submits the encrypted data to the provider. The storage provider:
    a. Uploads the data to the PAI Data BitTorrent network.
    b. Hosts the corresponding Torrent file.
    c. Generates a unique hash associated with the data.
    d. Returns the hash to Alice.
3. Alice creates a PAI Data transaction, with two outputs:

a. OP_RETURN transaction to Alice: This output contains the hash of the encrypted data in the OP_RETURN opcode field.
   b. Payment to provider (Optional): This output compensates the storage provider for hosting the Torrent file, based on an off-chain agreement among all parties.
4. Alice receives the OP_RETURN transaction and retrieves the data hash.
5. If and when Alice desires to make an ownership claim or otherwise retrieve the data, Alice: (1) retrieves the Torrent file from the storage provider using the data hash; (2) downloads the data from the PAI Data BitTorrent network; and (3) decrypts the data using her private key.

## Example: Sharing Transaction

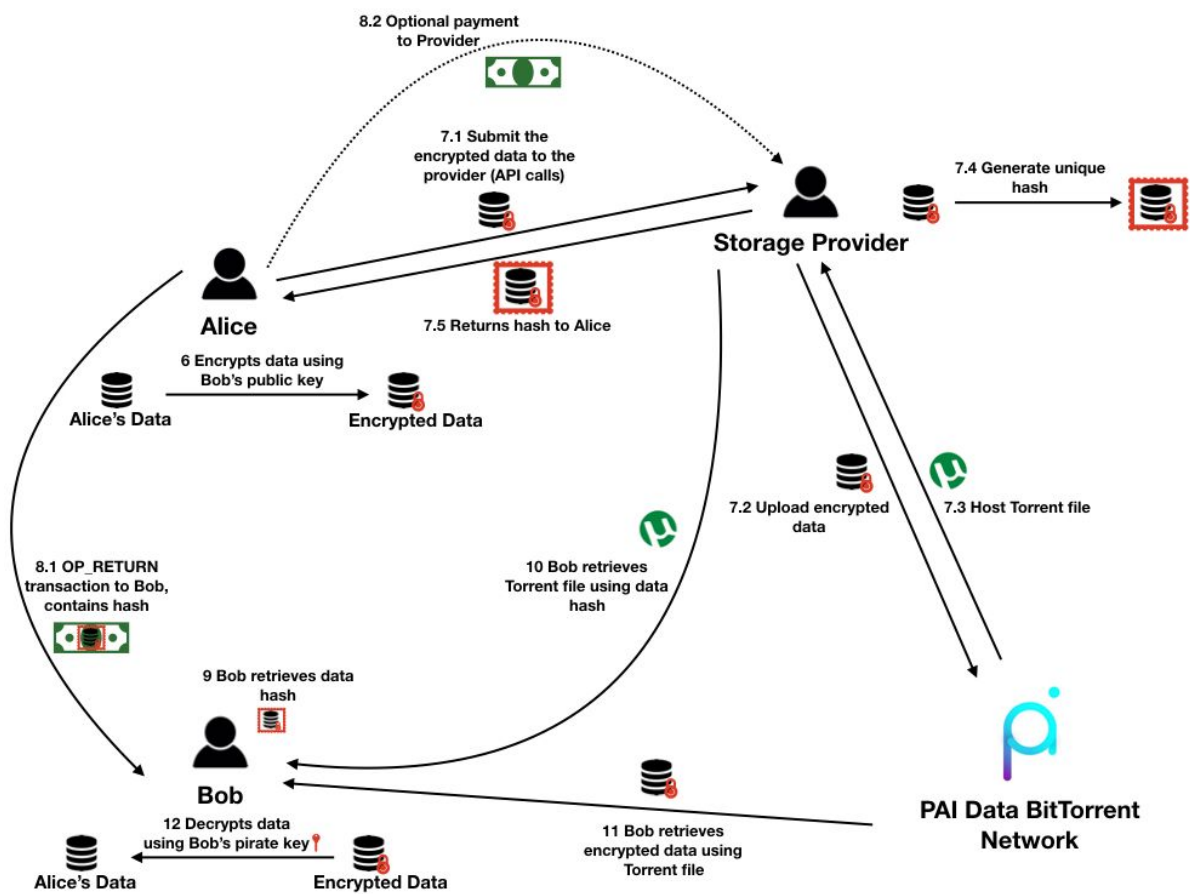

*Figure 2. Sharing Transaction Workflow*

As illustrated in Figure 2, two parties (e.g., Alice, the submitter, and Bob, the recipient) can share data with one another, via a PAI Data transaction, in the following way.

6. Alice encrypts the data with Bob's public key. This encrypted data can only be decrypted by Bob, who has the corresponding private key.

7. Alice chooses a storage provider and, through a series of standard API calls specified by the protocol (i.e., the "Provider API"), submits the encrypted data to the provider. The storage provider:
    a. Uploads the data to the PAI Data BitTorrent network.
    b. Hosts the corresponding Torrent file.
    c. Generates a unique hash associated with the data.
    d. Returns the hash to Alice.
8. Alice creates a PAI Data transaction, with two outputs:
    a. OP_RETURN transaction to Bob: This output contains the hash of the encrypted data in the OP_RETURN opcode field.
    b. Payment to provider (Optional): This output compensates the storage provider for hosting the Torrent file, based on an off-chain agreement among all parties.
9. Bob receives the OP_RETURN transaction and retrieves the data hash.
10. Bob retrieves the Torrent file from the storage provider using the data hash.
11. Bob downloads the data from the PAI Data BitTorrent network.
12. Bob decrypts the data using his private key.

## Comparison — Other Blockchain-Based File Storage Systems

Blockchain-based file storage systems have two main varieties [3]:

1. Proof of Storage / Capacity (e.g., Permacoin, Spacemint): Blocks are mined by proving the possession and integrity of some subset of the dataset that, as a whole, is too large for a single peer to store.
2. Mining a blockchain to broker between people offering storage and people wanting to buy storage, e.g., Storj, Filecoin (IPFS).

These alternative blockchain-based file storage systems typically modify the consensus mechanism of the protocol; PAI Coin does not. For example, the following consensus mechanisms are used in other solutions.

- Proof of Storage (e.g., Permacoin, Spacemint, Burstcoin, Chia): prove that the prover has reserved a certain amount of space.
- Proof of Replication (e.g., Filecoin/IPFS [4]): prove that some data has been replicated on one's uniquely dedicated physical storage.
- Proof of Spacetime (Filecoin/IPFS): prove that some data was being stored throughout a period of time.

# PAI Data's Advantages

## Lower barrier to entry for storage providers

In most other blockchain-based data storage solutions, the incentive for a node to provide storage is the block reward. The more storage or bandwidth a node can provide, the higher its probability to mine the next block and receive the block reward. This leads to competition in storage/bandwidth among nodes, just like the hash power competition in Proof of Work, and a node with little storage capacity may be gradually ruled out.

By contrast, in PAI Data, storage providers are optionally paid through transactions from data submitters. Individual storage providers without a large amount of available disk space could still benefit directly just from relaying data, provided an off-chain agreement is reached with the data submitter. This may help to lower the barrier to entry for storage providers, encouraging greater network participation. Grant and revoke operations are permanently embedded on the blockchain, serving as an immutable record of the chain of custody of the data, whether or not the data itself continues to be served by a provider.

## More flexibility in storage payment

The optional payment to the storage provider lends more flexibility in distributing fees. For example, the submitter is free to pay different fees to storage providers based on their accessible space, bandwidth, redundancy level and retention policy. Or, the submitter may pay a storage provider based on volume of data relayed. If data transmission is slow, from the view of recipient, the submitter may raise the fee they're willing to pay, to encourage more provider participation and increase speed.

## Higher efficiency in data transfer

Once a recipient obtains their data, and depending on the off-chain agreement between the submitter and recipient, the submitter may ask the storage provider(s) to delete their local copies through revoke calls. This can greatly improve system efficiency, allowing data providers to reuse their limited storage space for different data transmission tasks. Note that, in any case, data is encrypted with the recipient's public key, and is therefore unreadable by providers.